\documentclass{IEEEtran}

\newtheorem{thm}{Theorem}[section]

\usepackage{soul,color}   
\usepackage{epsfig}  
\usepackage{xspace}  

\newcommand{\name}{CollaPSE\xspace}
\newcommand{\myU}{myUnity\xspace}
\newcommand{\MyU}{MyUnity\xspace}
\newcommand{\mod}{ \mbox{ mod }}

\begin{document}
%
\title{Secured histories for presence systems}




%
\author{\IEEEauthorblockN{Eleanor Rieffel\IEEEauthorrefmark{1},
Jacob Biehl\IEEEauthorrefmark{1},
Bill van Melle\IEEEauthorrefmark{1}, and 
Adam J. Lee\IEEEauthorrefmark{2} \\
\IEEEauthorblockA{\IEEEauthorrefmark{1}FX Palo Alto Laboratory, \IEEEauthorrefmark{2}Univ. of Pittsburgh \\
Email: \{rieffel,biehl,billvm\}@fxpal.com}, adamlee@cs.pitt.edu \\ 
} 
}


\maketitle

\begin{abstract}

As sensors become ever more prevalent, more and more information will
be collected about each of us. 
A long-term research question is how best to support
beneficial uses 
while preserving individual privacy.
Presence systems are an emerging class of applications 
that support collaboration. 
These systems leverage pervasive sensors to estimate
end-user location, activities, and available communication channels.
Because such presence data are sensitive, to achieve 
wide-spread adoption, sharing models must reflect the
privacy and sharing preferences of the users.
To reflect users' collaborative relationships and sharing desires,
we introduce \name security, in which an
individual has full access to her own data, a third party 
processes the data without learning anything about the data values, 
and users higher up in the hierarchy learn only statistical information
about the employees under them. We describe simple schemes that 
efficiently realize \name security for time series data.
We implemented these protocols using readily 
available cryptographic functions, and integrated the protocols 
with FXPAL's \myU presence system.
\end{abstract}




\section{Introduction}

As sensors become ever more prevalent, more and more information will
be collected about each of us. This wealth of data has many benefits,
such as advancing medicine and public health, improving software and
services through user pattern analysis, and enabling each of us to
gain greater insight into our own habits and tendencies. At the same
time, the potential for misuse of such data is significant, and simply
the possibility that such data are being collected can ``lessen
opportunities for solitude and chill curiosity and self development''
\cite{Calo10}. A long-term research question is how best to support
beneficial uses while inhibiting less desirable effects.
For emerging classes of technologies such as presence
systems, addressing this concern is critical to adoption. 

\begin{figure}
\centering
\includegraphics[width=2.5in,height=4.0in]{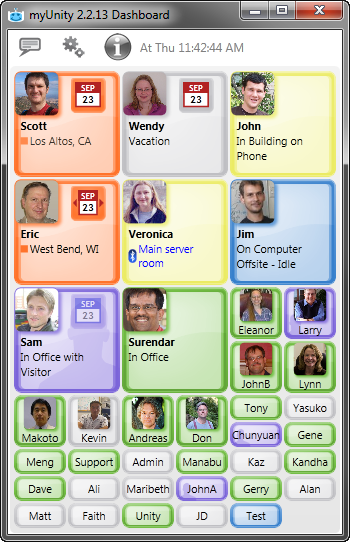}
\caption{\label{fig:MyUnityDashboard}\myU Dashboard.}
\end{figure}

Presence systems fuse physical sensing capabilities with social 
and communication software. Because sensor and 
presence data are sensitive, users' sharing preferences
must be considered by the designers of such systems, especially
for stored data.  FXPAL's \myU presence system
(Fig.~\ref{fig:MyUnityDashboard}), which has 
been in continuous use by more than 30 participants for over a year, 
was designed with these issues in mind. 
This paper describes the layer we added to \myU to address
many of our users' privacy concerns while enabling the benefits
that come with storage and analysis of presence data.

Feedback from \myU users indicates strong correlation between 
the extent to which a user is comfortable sharing data with a given 
person and how closely the user works with that person. 
Users were most comfortable sharing presence data with their closest 
collegues, expressed some comfort sharing data with their direct 
manager, and much less comfort with higher-level managers. 
We recently performed a formal survey that confirms these observations
more generally \cite{Biehl11}.
Our work supports an inverted hierarchical sharing structure that 
enables sharing of more detailed information with close colleagues, 
and less detailed information with people higher up in the hierarchy. 
This structure, which reflects collaboration relationships among users, 
is broadly applicable to social information sharing technologies.

In order to support collaboration in a wide variety of settings, 
presence systems must provide ubiquitous access to presence information. 
Therefore, a fundamental system design challenge is how to give users 
ubiquitous access on a variety of devices while also allowing them a high 
degree of control over and protection for their data.
\MyU users, for example, 
need access from mobile phones and tablets as well
as laptops and desktops. For this reason, data must be stored by
a third party, perhaps a server at their company or a cloud provider.
To preserve privacy, and to enable the users themselves to 
maintain full control of how their data are shared, the data must
be encrypted using keys controlled by the users and not shared
with the third party. 

In most instances, users are not interested in seeing raw historical data, 
only the trends derived from the data. 
A particularly useful case is when users are interested in
statistics for a group of people. A high-level manager may be
interested in statistics across all employees under her.
An employee may want to determine
times when people in the systems support group are less busy so
she can ask an involved but not urgent question. Users of \myU
and participants in our survey \cite{Biehl11}
expressed greater willingness to contribute their data
to group statistics than to share their individual values. 

To support these needs, we developed a mechanism to maintain 
confidentiality of user data while enabling contribution
to a statistic.
We propose \name (\underline{Colla}boration {\underline P}resence 
{\underline S}haring {\underline E}ncryption)
security in which, 
\begin{itemize}

\item at each time step, each member of a group encrypts her presence 
   values under her own key. Each individual has full access to her own data,

\item a third party that stores this encrypted data can compute encrypted 
  statistics, even over data encrypted under distinct user keys, without 
  learning anything about individual data values or the statistic computed, and

\item entities equipped with the appropriate keys can decrypt the group
  statistics without learning partial statistics or individual values.

\end{itemize}
We designed simple means to provide \name security 
for sums of time series data using
off-the-shelf cryptographic components efficient
enough to meet our real-time needs.

Because users are not always online, all of our protocols are 
non-interactive in that, after the initial setup, users
do not need to communicate with each other to encrypt or decrypt 
time series data. Moreover, to compute, the third party does not 
need to communicate with users (other than receiving 
the encrypted values). 
The protocols use a symmetric-key, additively homomorphic
encryption scheme \cite{Castelluccia09}. 
The more sophisticated protocols combine this 
encryption scheme with extensions of Chaum's DC-nets \cite{Chaum88} to
provide stronger privacy guarantees.
\name  security complements differential privacy, 
which limits what can be learned from the statistics.

The most significant contributions of this paper include:
\begin{itemize}

\item Definition of \name sharing structures that reflect the 
collaboration relationships among the participants.

\item Simple, non-interactive \name protocols for the
  case of sums over arbitrary subsets of time series data.



%
%

\end{itemize}


\section{Overview of \MyU}
\label{sec:MyUnityOverview}

The past few years have seen a rapid expansion of technologies 
that fuse physical sensing capabilities with social and communication 
software. One such system is \myU \cite{Biehl10}, a presence system 
for the workplace that supports collaboration by increasing workers' 
awareness of their colleagues' physical presence, 
activities, and preferred communication channels. 

 

\begin{figure}
\centering
\includegraphics[width=3.5in,height=2.0in]{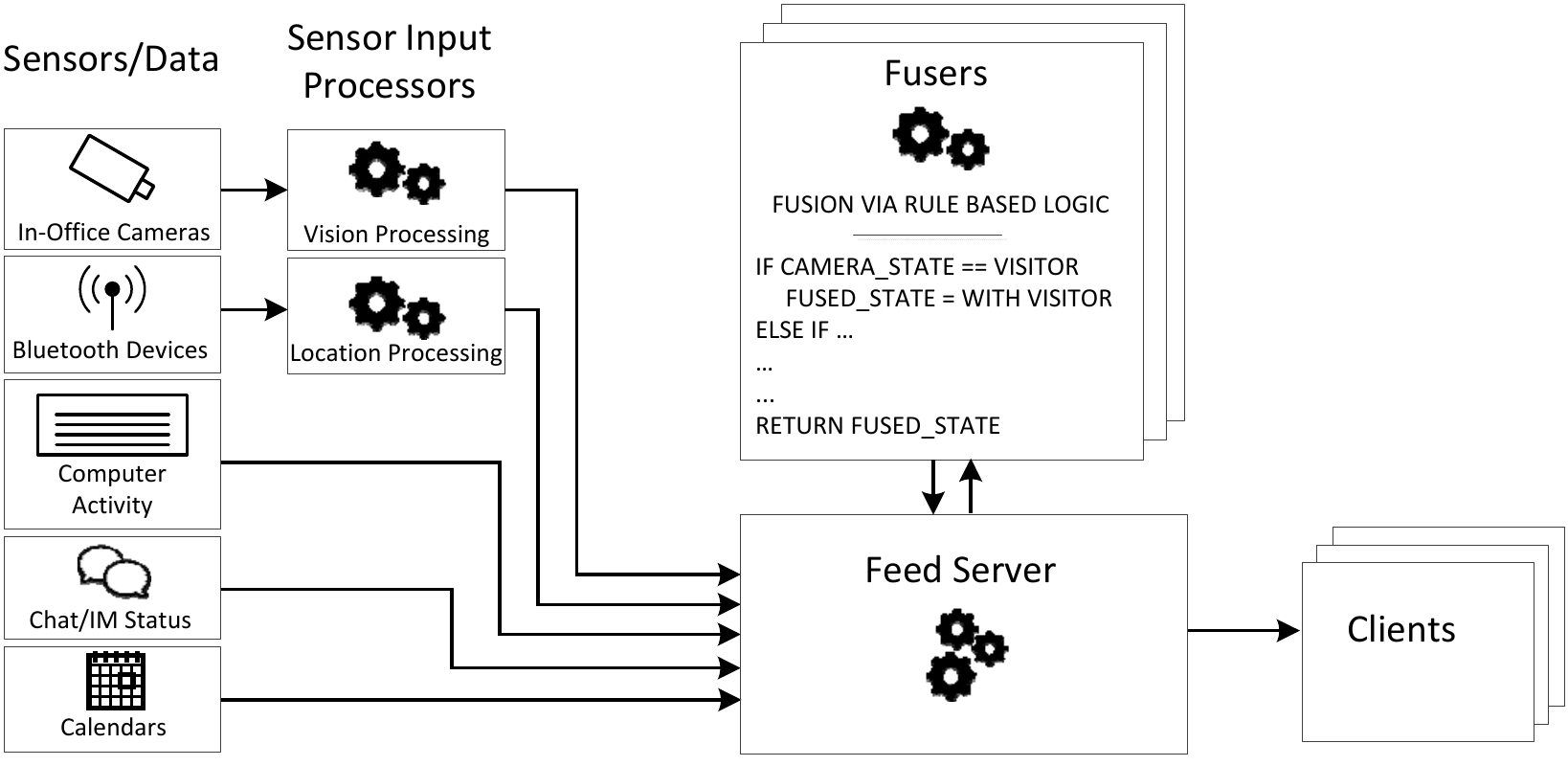}
\caption{\label{fig:MyUnityArch} Architectural overview of
the \myU presence system.}
\end{figure}

\MyU was designed to expand collaboration opportunities by
building group awareness.
\MyU collects data from cameras, bluetooth device sensors, 
mouse and keyboard
activity, network connectivity, IM availability, and the employee 
calendar (Fig.~\ref{fig:MyUnityArch}). 
At regular intervals, the data are aggregated and summarized 
into one of five presence states. 
A sixth state indicates there is insufficient data on the user. 
Users run clients that display presence states for colleagues 
as photo tiles within an awareness dashboard 
(Fig.~\ref{fig:MyUnityDashboard}). 
Each tile's color indicates the user's presence state: 
\begin{itemize}
\item {\bf Purple:} the person {\it has visitors} in her office.
\item {\bf Green:} the person is {\it in her office}. 
\item {\bf Yellow:} the person is {\it in the building}. 
\item {\bf Blue:} the person is {\it actively connected} remotely.
\item {\bf Orange:} the person is {\it connected via mobile} client.
\end{itemize}
The system represents each presence state as a five-bit string, 
in which each bit corresponds to one of the five positive presence states.
The six legal presence values are $10000$, $01000$, $00100$, $00010$, 
$00001$, and $00000$, corresponding to {\it in office}, {\it has visitor},
{\it in building}, {\it active online remotely}, {\it connected
via mobile client}, and {\it insufficient information}. 
The interface displays presence information for groups, such as  
the admin group, the support group, and the \myU research group,
as well as for individuals.

\MyU provides means for each participant to tailor which feeds 
she will allow; a user can turn off
any particular sensor feed whenever she likes. 
If, for instance, a user does not want a camera in her office,
this data feed can be left out, and a presence state can still be
computed. \MyU uses fusion rules that
adapt to missing information by degrading
the system's resolution of the user's state.  
When a colleague is visiting a user without a camera, for example,
the system will not report a visitor, but can report `In Office'  
if she is actively using her computer, or `In the Building' 
if she carries a detectable wireless device.

\MyU users are 
interested in sharing their presence histories or trend data with their
closest colleagues, but prefer that only aggregate statistics
are available to managers and employees outside their team.
This trust structure leads to the question of how to support a 
different type of hierarchical structure than is usually considered in the
access control literature, one in which higher levels in the hierarchy have
access only to summary statistics across a group of users, but
do not have access to data for individual users. 

To avoid concerns about misuse, the system initially did not store 
any data.  Users consistently expressed interest, however, in 
seeing personal trends, activity
patterns of coworkers, and data pooled across groups of users. 
They gave many examples of how access to  historical data 
would support collaboration, such as knowing
when a colleague usually returns from lunch on a Friday or whether
the support team tends to have many visitors Monday morning.
At the same time, 
users expressed concern about misuse if data were
stored, and a strong desire for complete control over any stored
data. 
Many were
willing to contribute data to statistical analyses 
so that the designers could analyze the usage of the system or
other users could get statistical information about a
group as a whole. 


\MyU has been well received by its users, who have
incorporated it into their daily routine to help 
coordinate collaboration with colleagues. A field study \cite{Biehl10}
showed an increase in face-to-face communication, most users'
preferred means of communication, after adoption of \myU.
Nearly all users have continued using the system after the trial
run. The popular press, while recognizing \myU's benefits
in supporting collaboration, has presented a creepy view of the 
system with headlines such as ``Someone's watching you" \cite{Simonite10}.
Such a viewpoint illustrates that to achieve widespread adoption,
presence systems must address user privacy concerns. 

\section{Secured histories}
\label{sec:SecHist}

Feedback from users indicates significant value in storing 
historical data, but only if secured and equipped with an appropriate 
sharing structure. This section defines the problem more 
precisely and describes components used in our solutions.


\subsection{Problem definition}
\label{sec:probDef}

We state more precisely the requirements for
\name security: 
(i) at each time step, each user encrypts her own data under her own key, 
(ii) a third party can compute encryptions of sums over
arbitrary subsets of a user's data without learning anything about 
the values,
(iii) the third party can compute encryptions of sums over data contributed
by multiple users encrypted under different keys, and
(iv) users with the appropriate set of keys can decrypt a sum without
learning anything about the contributing values other than what
can be deduced from the sum.

We formalize the problem in terms of the following players:
\begin{itemize}
\item $n$ {\it team members}, each of whom has a value, such as a presence 
state, to contribute at each time step, 
\item an {\it analyst} who wishes to obtain a statistic over these values, and
\item  an honest-but-curious {\it third party} who contributes to
the computation without learning anything about the values.
\end{itemize}
There may be one or more analysts. 
Analysts may be managers or may be one or more of the team members. 
We use the term in the formal definition so as not to prejudice
which entities have those capabilities. A user may be
an analyst for one group and a team member of another.

We define \name protocols for sums over time series data. From sums,
many important statistics can be determined. To obtain
the average, the user divides by the number of terms, which
the user may know, or may be supplied by the third party.
Because presence states in our case are Boolean values, the
variance can be computed directly from the average: $V = A - A^2$.
In a non-Boolean case, the square of each value $v_i^2$ can be
encrypted and stored, and the decryption of the sum of such values,
together with the average, gives the variance.

A \name security protocol for sums with respect to
time series data contains the following algorithms: \newline
{\bf Setup:} Establishes public parameters and constants used 
by all parties in the protocol. \newline
{\bf Generate Keys:} Establishes the key structure.
It is run once, prior to any of the data generation time steps. \newline
{\bf Encrypt:} At each time step, each individual encrypts her
values under her own key or keys and sends the encrypted values to 
the third party for storage.\newline
{\bf Compute Encrypted Sum:} The third party can compute an
encryption of the sum over any specified set of data.\newline
{\bf Decrypt Individual Sum:} Any individual with access to
individual $A$'s keys can decrypt the sum over an arbitrary subset
of individual $A$'s values.\newline
{\bf Decrypt Group Sum:} An analyst with the appropriate set of keys 
can decrypt the sum over all values, at a given time step,
for a group of users.
As we will see in Section \ref{sec:accessCtrl}, the ``appropriate 
set of keys'' with which the analyst can decrypt varies from
protocol to protocol, as does the key structure.


A \name protocol is {\it secure} if {\it (i)}
an honest-but-curious third party can learn
nothing about the data values, {\it (ii)}
an analyst learns nothing about individual
users' values other than what she can deduce
from the statistics, and {\it (iii)}
each user learns
nothing about other users' values. 

A \name protocol for time-series data is {\it non-interactive} if, 
after the setup phase, the users do not need to communicate with each other, 
and each user only communicates with the third party to deliver
the encrypted data at each time step.

A \name protocol is {\it secure against $k$-collusion}
if for any set of $k$ or fewer
parties, whether consisting of team members, outsiders, or analysts, 
the colluding group cannot learn anything about another person's data 
other than what can be deduced from the colluding members' data and
the full statistic (if an analyst is part of the colluding group).

\subsection{Secured histories architecture}
\label{sec:arch}

\begin{figure*}
\centering
\includegraphics[width=7.4in,height=3.0in]{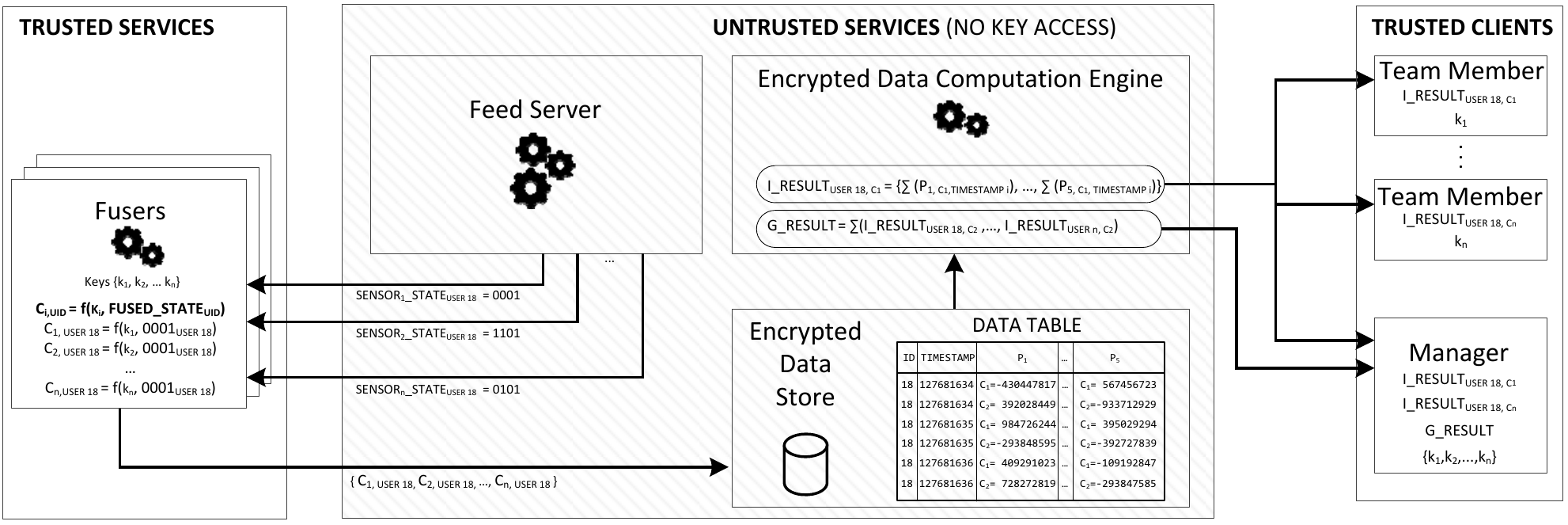}
\caption{\label{fig:SecHistBasic} Basic architecture for
the Secured Histories system.}
\end{figure*}

The system architecture 
(Figs.~\ref{fig:MyUnityArch}, \ref{fig:SecHistBasic}) includes 
raw data sources, such as cameras, bluetooth device sensors, 
and keyboard monitors. These send their data, along
with metadata, such as source ID and timestamp, to sensor-input 
processors that process the data and send it
to the Feed Server. A video feed processor, for example, 
takes in raw video streams, but sends to
the Feed Server only compact descriptions of events observed.
In some cases, raw data sources may talk directly to the Feed Server.
The Feed Server forwards data to the appropriate fuser, which
computes the presence states. 

The following components play a role in our protocols:

{\bf TRUSTED (partial access to keys)} \newline
{\bf Fusers:} 
There is one fuser per individual. It has access to the keys
used to encrypt its individual's data. 
It computes its individual's presence state from data received 
from the Feed Server, encrypts this presence state using the
individual's keys, and returns the
encrypted presence state to the Feed Server, which routes it to
the Encrypted Data Store.\newline
{\bf Client:} 
A given individual may run multiple clients on different desktops or
mobile devices. Each client has
access to that individual's keys, and the keys for any other individuals 
who wish to share their historical presence data with that individual. 
Clients decrypt and present information in the client interface,
and pass user queries to the Feed Server. 

{\bf UNTRUSTED (no access to keys)} \newline
{\bf Encrypted Data Computation Engine:} The Encrypted Data
Computation Engine computes on encrypted data and 
returns the results to the Feed Server to be sent to the clients. \newline
{\bf Encrypted Data Store:} The Encrypted Data Store
stores the encrypted data, together with its metadata. 
It also keeps a list of missing data
ranges. When the store returns aggregated results, it includes a list,
often empty, of any missing data. \newline
{\bf Feed Server:} 
The Feed Server routes information between the various components 
of the system.  

Instead of having one fuser per individual, 
members of a team who trust each other could share a fuser.
The untrusted components could reside in a public cloud. 
More than one of each of the untrusted components may
be needed to support a large organization.




\subsection{Underlying encryption scheme}
\label{sec:background}


To meet property (i) of the problem definition, 
any efficient encryption scheme could be used.
Presence states are Boolean values, so schemes that encrypt
Boolean values compactly will support more efficient storage and
transmission. 
To meet (ii), any additively homomorphic encryption scheme 
can be used. Most homomorphic encryption schemes are
public key schemes that do not encrypt Boolean values compactly.
We selected Castelluccia et al.'s symmetric-key based scheme 
\cite{Castelluccia09} in part because of its compact and efficient encryption.
Property (iii) is more challenging to meet, because most existing
homomorphic encryption schemes do not support combining values that 
have been encrypted under distinct keys. 
Castelluccia's scheme does support homomorphic 
addition of values encrypted under distinct keys. 
To obtain property (iv), we devised a complex key structure with
which to augment Castelluccia's scheme.


In Castelluccia et al.'s cryptosystem, values are encrypted
by adding a {\it pad}, obtained from a 
pseudorandom function and a nonce $n_t$, mod $M$, and decrypted by
subtracting it.  More specifically, in our setting, let $X_i$ denote
a user, where $i$ is an index over the user population.
Individual $X_i$ with key $k_i$ encrypts value $v_i$ at time $t$ by
evaluating a pseudorandom function $g_{k_i}$ at nonce $n_t$ 
and adding it to $v_i$ to obtain
$$c_i = v_i + g_{k_i}(n_t) \mod M.$$ 
To decrypt, she computes $g_{k_i}(n_t)$ and subtracts it from $c_i$:
$v_i = c_i - g_{k_i}(n_t) \mod M$.

This cryptosystem is parametrized by a pseudorandom function (PRF) family,
a collection 
$F_\lambda = \{f_s: \{0,1\}^\lambda\to \{0,1\}^\lambda\}$ 
of functions indexed by security parameter $\lambda$.  
Since provably secure pseudorandom functions are very slow, 
Castelluccia et al.~\cite{Castelluccia09} advocate using keyed hash 
functions such as HMAC 
followed by a
length-matching hash function $h$ that does not need to be 
collision-resistant, but must have uniform output upon uniform input.
The simple hash 
function $h:\{0,1\}^\lambda \to \{0,1\}^\mu$
that partitions the $\lambda$-bit output of $f_s$ into length
$\mu$ substrings and adds them together is an example of such a function. 
Applying such a function $h$ ensures that if at least one
of the blocks is indistinguishable from random, then the output of
the composition of $h$ with $f_s$ is indistinguishable from random.
Applying $h$ is unnecessary with a provably secure pseudorandom function.
In~\cite{Castelluccia09}, the authors prove this scheme semantically
secure:
\indent{
\begin{thm}
  Assuming $F_\lambda = \{f_s : \{0,1\}^\lambda \rightarrow
  \{0,1\}^\lambda\}$ with ${s \in \{0,1\}^\lambda}$ is a PRF,
  and $h : \{0,1\}^\lambda \rightarrow \{0,1\}^l$ satisfies 
  $\{t \leftarrow \{0,1\}^\lambda : h(t)\}$ is uniformly
  distributed over $\{0,1\}^l$, the above construction is semantically
  secure.
\end{thm}
}
The simple $h$ above satisfies the uniformity condition, 
so the security reduces to that of the PRF used. HMAC is a PRF provided 
the underlying compression function is a PRF \cite{Bellare06}. 

This cryptosystem provides the ability to combine values 
homomorphically that are encrypted under the same or different keys.  
Consider individuals $X_1$ and $X_2$ with keys 
$k_1$ and $k_2$, respectively.
They wish to encrypt the values $v_1$
and $v_2$, respectively, at time $t$.  Each encrypts by evaluating
her pseudorandom function $g_{k_i} = h(f_{k_i})$ indexed by $k_i$ at $n_t$:
\begin{eqnarray*}
c_1 &=& v_1 + g_{k_1}(n_t) \mod M \\
c_2 &=& v_2 + g_{k_2}(n_t) \mod M.
\end{eqnarray*}
Given the aggregate ciphertext $c = c_1 + c_2$, an individual with access to
both $k_1$ and $k_2$ can construct the sum
$r = g_{k_1}(t) + g_{k_2}(t)$ and recover the aggregate value 
$$v = v_1 + v_2 = c - r \mod M.$$ 


\section{The base protocol}
\label{basicScheme}

This section describes a non-interactive protocol for sums over time 
series data that satisfies all of the conditions of \name security
except that an analyst can decrypt the 
individuals' values. Section \ref{sec:accessCtrl} extends this
protocol to obtain full \name security, in which the analyst can 
decrypt only the sum, not any of the individual values.

\subsection{Base protocol description}

Each fuser computes, at regular intervals, the current presence
state for its user and sends it to the Feed Server to send to clients.  
It encrypts each bit of a five-bit presence string separately in 
order to support computation of statistics restricted to one
type of presence state.
The fuser encrypts with the user's key, taking the timestamp concatenated 
with the presence state type 
as the nonce. 
At each time step, the fuser sends a record, consisting of
a user ID and timestamp, both unencrypted, and five encrypted Boolean 
values, one for each presence type, to the Feed Server to be
placed in the Encrypted Data Store.  \newline
{\bf Setup:} {\bf (i)} Establish a modulus $M$ large enough
for the application at hand. 
The modulus must be larger than the number of terms that would 
ever contribute to the computation of a single statistic. 
The bit length of encrypted values will be 
$\mu = \lceil \log_2(M) \rceil$.
{\bf (ii)} Establish a pseudorandom function family 
$F_\lambda = \{f_s:\{0,1\}^\lambda \to \{0,1\}^\lambda\}$, and choose 
$\lambda$ according to the desired level of security. 
{\bf (iii)} Establish a length-matching hash function 
$h: \{0,1\}^\lambda \to \{0,1\}^\mu$.  \newline
{\bf Generate Keys:} Each individual $X_i$ runs a key generation
algorithm to obtain a key $k_i$.  \newline
{\bf Encrypt:} At each time step, each individual $X_i$ encrypts 
each of the five 
bits $m_j$, for $j = 1, \dots, 5$, of a presence state $m$ as 
$$c_j = m_j + h(f_{k_i}(n_j)) \mod M,$$ 
where the nonce $n_j$ is 
the concatenation of the presence state type and the
timestamp. We refer to $r_j = h(f_{k_i}(n_j))$ as a {\it pad}. 
The record that is transmitted includes a header, containing the user ID
and timestamp transmitted in the clear, followed by the five
ciphertexts $c_j$.  \newline
{\bf Compute Encrypted Sum:} 
The Encrypted Data Computation Engine adds ciphertexts mod $M$ 
to obtain a ciphertext sum $c$. 
\newline
{\bf Decrypt Group Sum:} A user with access to the keys for all users
whose values contribute to a sum can decrypt an encrypted sum $c$ 
by computing pads for all contributing values and subtracting them
from $c$ mod $M$. 

Sections \ref {sec:ind} and \ref{sec:basicGpStats} give example decryptions 
of a sum.
In order to decrypt a sum, a user with access to the appropriate keys must
also have access to the appropriate nonces. 
Because data are collected at regular
intervals, users know which timestamps should contribute
to the sum. 
In order to handle missing data, the Encrypted Data Computation
Engine sends the client a list of any expected triples 
(timestamp, user ID, presence type) that are missing
from the sum. Since the system is robust, 
usually this list will be empty or very small. 

\subsection{Example: queries about an individual}
\label{sec:ind}

A user can query the Encrypted Data Store about her own history, 
receiving encrypted values that she
can decrypt using her key. 
She can also query the Encrypted Data Computation Engine to
receive encrypted sums. For example, she may want to
understand her typical daily presence pattern by dividing the day into
fifteen-minute intervals and requesting the totals of each type of presence
state for each fifteen-minute interval over the past three weeks.  The
Encrypted Data Computation Engine computes and returns encrypted 
sums for each type of state in each interval. 
She then decrypts each encrypted sum using her
key and the nonces. The semantic security of the cryptographic
construction used to encode each presence state ensures that the
Encrypted Data Computation Engine cannot learn any information about 
her presence states.

Instead of estimating her presence state pattern from the data over the 
last three weeks, she may wish to use data from the past six months,
but weight the more recent data more heavily. After receiving the 
encrypted weighted sum from the Encrypted Data Computation Engine, 
she decrypts using the same weighting to sum the pads. 
As a simple example, suppose she wants to obtain the weighted sum 
$v = v_1+2v_2$, where $v_1$ is the sum over the earlier data, and $v_2$ the 
sum over the recent data. She asks the Encrypted Data Computation Engine 
to compute $c = c_1 + 2c_2$ mod $M$. Knowing that $c_1 = v_1+r_1$ and 
$c_2 = v_2+r_2$, 
she can decrypt by subtracting from $c$ the similarly weighted sum of 
the pads, $r = r_1 + 2r_2$ to obtain
$$v = c_1 + 2c_2 - r_1 -2r_2 \mod M.$$


\subsection{Example: queries about a group of users}
\label{sec:basicGpStats}


Suppose each of $L$ team members sends her key to an analyst.
At a given time, and for a given presence type, all team members' 
values are encrypted using the same nonce $n$, a concatenation 
of the timestamp and the presence type.
Each fuser encrypts its team member $X_i$'s 
value $v_i$ by adding the pad $r_i = h(f_{k_i}(n))$ to $v_i$  
modulo $M$:
$c_i = v_i + r_i \mod M$. 
The analyst can request the sum from the Encrypted Data Computation Engine,
which is
$$c = \sum_{i = 1}^L v_i +  \sum_{i = 1}^L r_i \mod M.$$ 
The analyst can compute the pads $r_i$ since she has
all of the keys and knows all of the nonces. She can even compute
the sum of the pads prior to receiving $c$ from the Encrypted 
Data Computation Engine. 
She subtracts this sum, $\sum_{i = 1}^L r_i$ from $c$
to obtain the total $v = \sum_{i = 1}^L v_i$. 

Advantages of this approach over 
having the client perform the computation after receiving,
decrypting, and summing the contributing values include
(i) more efficient bandwidth use, 
and (ii) improved security, in that raw presence values are not seen
in decrypted form. 

The amount of computation required to decrypt a group statistic scales 
linearly with the number of values contributing to the statistic, since 
the computation of the pads forms the bulk of the computation. 
The computation of these pads can be computed prior
to receiving the encrypted value, so the part of the decryption
that must take place after receiving an encrypted sum is constant:
only one value, the sum of the pads, must be subtracted to decrypt,
and this subtraction is much faster than a single decryption by
a public key homomorphic encryption scheme.
For this reason, comparison of decryption times for sums 
between our protocol and public key based
homomorphic encryption schemes is not straightforward. 
For large sums, the computation of the pad sum is expensive, but can 
be computed ahead of time, prior to receiving the encrypted sum. 
In contrast, the decryption time for public key homomorphic encryption 
schemes is constant, no matter how many terms contribute to the sum,
but decryption can start only after the encrypted sum has been received. 
In the extensions of this protocol given in Section \ref{sec:accessCtrl},
the cost of decrypting group sums does not increase
with the number of users.

\subsection{Application of the base protocol}
\label{sec:implAndEval}

\begin{figure}
\centering
\includegraphics[width=3.0in,height=2.5in]{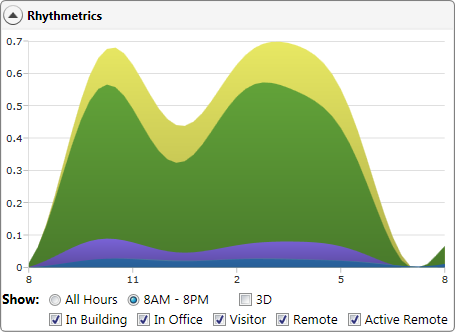}
\caption{\label{fig:historiesGraph}Graph summarizing an individual's
activity history.}
\end{figure}

Our initial implementation supports the computation of a rough summary
of a single user's presence pattern from encrypted stored data. 
To obtain baseline efficiency estimates, we used
presence data for one individual from a roughly three-week period.
This test set consists of $31,568$ records, collected once a minute, 
each with five encrypted values, for a total
of $157,840$ encrypted values. 
For the statistical summary of this person's daily presence pattern, 
we aggregated the presence states over fifteen-minute intervals, 
and summed over the $21$ days of data, to
obtain histograms for each of the five presence states. 
We smoothed
to further obscure the data and make it more visually appealing. 
Fig.~\ref{fig:historiesGraph} shows the graph our system produced.
The colors are the same ones used on the tiles in 
Fig.~\ref{fig:MyUnityDashboard}) to indicate the presence states. 

We implemented the core functionality in Java. 
We ported some of this code to our .NET clients.
We used HMAC 
as implemented in javax.crypto and .NET with default security
parameter $\lambda = 128$. 
We wrote a length-matching hash function that splits a byte array
into groups of four bytes and adds these together. 
For convenience, we took $M = 2^{32}$, 
so that each encrypted value is $32$ bits,
but we could have used a considerably smaller modulus.
The bit-length of the encrypted data is an order of
magnitude smaller than that needed by a public key homomorphic
encryption scheme with a similar level of security. Thus, our
protocol has more efficient storage and bandwidth usage than
public key solutions.

 
We benchmarked our protocol on
a virtualized Windows Server 2008 instance, hosted
by a Citrix XenServer hypervisor, which was allocated four virtual
CPUs with $8$ GB of memory, an $80$ GB virtual disk, and a $1$ GB
full duplex ethernet port. The underlying Intel Xeon E5450 
hypervisor CPU runs at $3.21$ GHz.  Our clients vary, but our numbers 
are from an Intel $2.40$ GHz dual core with $2$ GB of RAM. 
We made no attempt to optimize the code. 
On our server, each encryption took roughly $2.33$ milliseconds. 
Computation of all $480$ sums,
$96$ fifteen-minute intervals per day for each of the five presence states,
took $439$ milliseconds, or about $0.92$ milliseconds per sum 
with approximately $2105$ contributing values. 
Computing the pads for decrypting all $480$ sums is slow,
taking $11.5$ seconds total, but these pads can be computed 
prior to receiving the encrypted sum.  
The final decryption takes $2.33$ milliseconds per sum, 
or $1.12$ seconds for all sums contributing to the graph. 


\section{Secured histories: \name protocols} 
\label{sec:accessCtrl}

The basic protocol 
of Section \ref{basicScheme} enables an individual to use
an honest-but-curious third party to store and 
compute on her data. The protocol enables a fully trusted analyst 
who has access to all keys to use the third party to aid in computing 
sums over values from multiple individuals that have been 
encrypted under different keys. 
This section extends the basic protocol to a series of 
increasingly sophisticated non-interactive protocols in which
an analyst can decrypt only the sum, but not the individual values 
or any sub-sum.  The more sophisticated schemes
guard against $k$-collusion.
As a side benefit, decryption of group sums
is faster than in the basic protocol: the time for decrypting a sum
in the protocol of Section \ref{sec:twoLevel} is constant, whereas
in the basic protocol, it increases
linearly with the number of values contributing to the sum.
These schemes can be nested to support an inverted hierarchical 
sharing structure in which
nodes at higher levels can decrypt sums over
all nodes below them, but cannot decrypt any partial sums,
including individual values.

\subsection{A \name protocol}
\label{sec:twoLevel}

\begin{figure}
\centering
\includegraphics[width=3.3in,height=2.2in]{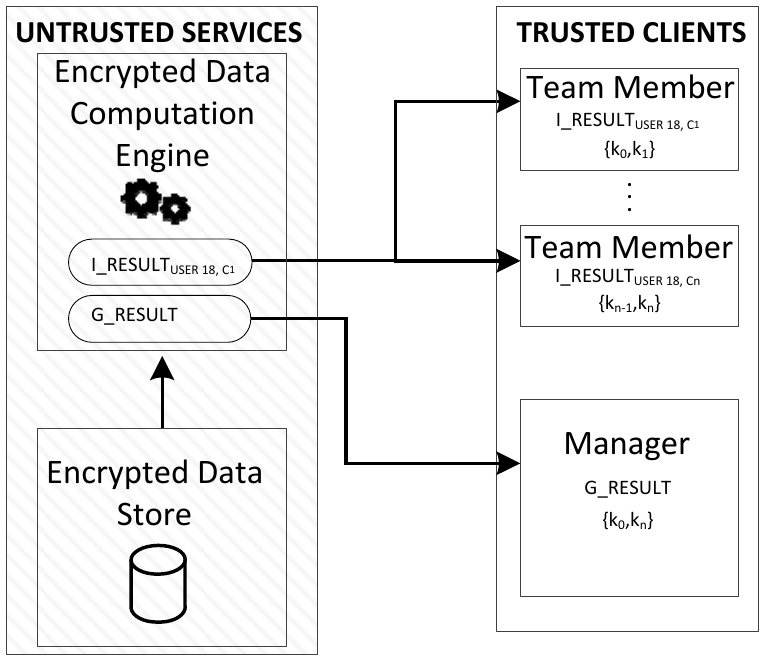}
\caption{\label{fig:SecHistComplex} Secured histories architecture 
with key assignments for the \name protocol of Section \ref{sec:twoLevel}.}
\end{figure}

Suppose a project team wants a manager
to see only pooled data on the team's activities.
The manager may see the pattern of availability of the group, 
for example, without learning anything about
the pattern of any individual, other than what
can be deduced from the statistics for the whole group. 
We describe a non-interactive protocol in which $N$ 
team members $X_i$ encrypt one value at each time step 
in such a way that the manager
can decrypt the sum but not the individual values. 

Ideally we would solve this problem by providing the manager with 
a key $k$ and the team members with keys $k_1, \dots, k_n$ such
that at each time step, and for each presence state, the 
pad computed from the manager's key $k$ is the sum of the pads 
computed from the team members keys  $k_1, \dots, k_n$. We are
not aware of a method for obtaining $n$ pseudorandom functions
$g_1, \dots, g_N$ and another pseudorandom function $f$ such
that $f(x) = \sum g_i(x)$ for all  $x$, with the property
that the ability to compute $f$ does not confer the ability to
compute any $g_i$. We take a 
less direct approach, using an extension of Chaum's DC-nets~\cite{Chaum88}.
Whether there is a more direct approach is an intriguing open problem.

The following algorithms constitute a \name
protocol for sums with respect to time series data: \newline
{\bf Setup:} 
Same as for the base protocol (Section \ref{basicScheme}). \newline
{\bf Generate and Share Keys:} 
Each team member $X_i$, for $1 \leq i < N$, generates a key $k_i$, and
the manager generates a key $k_0$.
Each individual member $X_i$ sends her key to individual $X_{i+1}$ where 
the indexing is modulo $N+1$ and the manager is considered
individual $X_0$. \newline
{\bf Encrypt:} 
Each team member $X_i$ encrypts her value $v_i$ at time $t$ by 
adding the pad $r_{i-1} = h(f_{k_{i-1}}(n_t))$ and then 
subtracting the pad $r_i = h(f_{k_i}(n_t))$ from $v_i$, 
$$c_i = v_i + r_{i-1} - r_i \mod M.$$ 
All team members 
use the same nonce $n_t$, a concatenation of the timestamp
with the presence state type.\newline
{\bf Compute Encrypted Sum:} The third party can compute an
encryption of a sum over any specified set of data.\newline
{\bf Decrypt Individual Sum:} Anyone with access to
individual $A$'s keys can decrypt sums over arbitrary subsets
of individual $A$'s values.\newline
{\bf Decrypt Group Sum:} Anyone (e.g. the manager) with the two keys 
$k_0$ and $k_N$ can decrypt sums over all values, at a given time step,
for the whole team.
When the encrypted values $c_i$ at time $t$ are summed over the group,
all of the pads cancel except for $r_0$ and $r_N$:
$$c = \sum_{i=1}^N c_i = v_1 + \dots + v_N + r_0 - r_N \mod M.$$ 
Anyone with keys $k_0$ and $k_N$ can compute
pads $r_0$ and $r_N$ to decrypt $c$ to obtain the sum $v = \sum v_i$.
Sums over the group's values at multiple times can be similiarly decrypted.

All players have two keys.
The manager has keys $k_0$ and $k_N$, and 
each team member $X_i$ has keys, $k_i$ and $k_{i-1}$, 
as shown in Fig.~\ref{fig:SecHistComplex}.
The key structure is a chain in which pads computed from 
the keys cancel in the desired way in a sum.
Because the manager does not have any of the other keys, 
she cannot decrypt any subtotal, let alone any individual value $v_i$. 
To decrypt the sum, she needs to compute only two pads; thus 
decryption of group sums is more efficient for
this protocol than for the basic protocol of Section \ref{sec:basicGpStats}.

A team may choose to give other outsiders the same keys as the manager, 
in which case multiple people can decrypt the statistic, but not
the individual values.
This example illustrates that the hierarchical
structure of the keys does not necessarily correspond directly with the 
access control structure, which is determined by who receives which
keys.

\subsection{A family of \name protocols}

In the protocol of Section \ref{sec:twoLevel}, a manager cannot 
decrypt individual team members' values, but two players can collude to 
decrypt another's data. 
Players $X_{i-1}$ and $X_{i+1}$ can together decrypt team 
member $X_i$'s value, where the manager is 
team member $X_0$, and the indexing is mod $N+1$.
We can guard against $s$-collusion by increasing to $s+1$ the
number of pads used to encrypt each value and distributing the
keys in such a way that all pads except the
manager's cancel in the sum, and no subset of the $s$ players
knows enough keys to decrypt another member's value.  \newline
{\bf Setup:} Same as for the protocol of Section \ref{sec:twoLevel},
with the addition of a graph structure in which every team member and the
manager has exactly $s+1$ neighbors, for $s$ odd. \newline
{\bf Generate and Share Keys:} 
Team member $X_i$ generates keys $k_{ij}$ for every $j < i$ such that 
$X_j$ and $X_i$ are neighbors.
Team member $X_i$ shares $k_{ij}$ with neighbor $X_j$. \newline
{\bf Encrypt:} 
Each team member $X_i$ encrypts her value $v_i$ at time $t$ by 
adding pad $r_{ij} = h(f_{k_{ij}}(n_t))$ for every neighbor $X_j$
with $j<i$ and subtracting the pad 
$r_{ij}$ for every neighbor $X_j$ with $j>i$,
where all arithmetic is done modulo $M$:
$$c_i = v_i + \sum_{j\in nbhr(i)} (-1)^{\chi_i(j)}r_{ij} \mod M,$$ 
where $\chi_i(j)$ is $0$ for $j<i$ and $1$ for $j>i$. \newline
{\bf Compute Encrypted Sum:} Same as before.\newline
{\bf Decrypt Group Sum:} In a sum of ciphertexts over all
team members at a particularly time, all pads cancel
except for the $r_{ij}$ with $j = 0$. Since manager $X_0$
has keys $k_{i0}$, she can decrypt sums over the whole team.

These protocols generalize to support multi-level inverted hierarchies 
in which nodes at higher levels can decrypt
only summary statistics over all leaf nodes below them, and cannot 
decrypt lower-level statistics or values.

\section{Related work}
\label{sec:relWork}

Several commercial and research systems support awareness in organizations.  
Most provide awareness of a single channel of information.
In systems such as Portholes \cite{Dourish92},
workers observe the activity of co-workers via video feed.
Fogarty and Hudson's toolkit \cite{Fogarty07}
used computer activity, ambient sound, and other sensors,
to predict a person's level of interruptibility. 
Other systems (e.g., \cite{Begole02,Danninger06, Fogarty04,Horvitz03})
performed similar functions with different configurations of
sensors. 
Most of these systems do not save past state, and none have 
adequate mechanisms for protecting and controlling access to historical
data. Shared calendars sometimes provide control over 
how long data are retained and how historical 
information is accessed, and chat clients often provide 
user control over whether chat logs are retained.
While we applied our sharing scheme to the \myU system, 
our work could be adapted to work with other tools.



In Castelluccia et al.'s \cite{Castelluccia09} setting,
data aggregation in wireless sensor networks,
no data are stored, and a fixed computation is carried out as
the data traverse the network. 
They have essentially one client, whereas we have many.
To the best of our knowledge, our work is the
first application of their symmetric homomorphic
encryption scheme outside of the wireless sensor network area.

Our approach differs from secure multi-party computation (SMC) in
a number of respects. Our approach is non-interactive in that,
apart from key sharing, which is done once prior to any data
storage, and is only updated when sharing relationships change. 
After that, unlike in SMC approaches, all statistics are computed 
without the team members communicating with each other. All 
computation is done by the third party, not the team members, who 
do not even need to know what statistics are being computed. 
Our approach is not restricted to a single round of data,
but rather handles time series data of arbitrary length without
requiring further key generation or key sharing.
Furthermore, our approach is substantially more efficient than general 
SMC constructions.

Molina et al.~\cite{Molina09} 
study how to enable clinical research without giving 
patient records to the researchers. In their solution, caregivers,
who have full access to patient records, use
multiparty computation with public key homomorphic encryption to 
answer researcher aggregation queries.


Differential privacy foils deduction of individual attributes from data
such as aggregate statistics, a concern complementary to our own.
In the standard setting, the differential privacy mechanism is
carried out by a trusted curator who has access to all data.
Rastogi and Nath \cite{Rastogi10} provide differentially private
aggregation of encrypted data using Paillier threshold homomorphic
encryption to achieve differentially private aggregation without a
trusted curator. Their decryption, unlike ours, is multiparty. 

 
\section{Discussion}
\label{sec:disc}

While the protocols of Section \ref{sec:accessCtrl} 
were designed to support full \name security,
they have the added benefit of reducing to a constant
the number of pad computations needed to decrypt.
For this reason, in large organizations in which statistics are desired
over a large number of employees, it may be worth implementing one of the 
more sophisticated protocols of Section \ref{sec:accessCtrl} 
even if an analyst is given all keys. 
 
A number of residual risks remain. We were not concerned with 
protecting the integrity of the data. Because the encryption is
homomorphic, it is malleable, so a separate mechanism, such as
the one Castelluccia et al.~\cite{Castelluccia09} provide, is needed
to protect against tampering with encrypted records.
A more significant risk is that, from the release of aggregate 
statistics, individual values could be deduced.
As mentioned in Section \ref{sec:relWork}, 
differential privacy mechanisms address this threat.
The simple structure of our schemes means that they can be combined
with differential privacy techniques to support the computation of
statistics with a differential privacy guarantee without the need for
the individual contributors to share their individual values with
anyone, including a curator. 

To increase the minimum number of users who can 
successfully collude to decrypt another user's values, 
the protocols of Section \ref{sec:accessCtrl}
require each user to use more keys to encrypt each value. 
An interesting open problem is how to support a structure in
which a manager can decrypt an aggregate of all team members values,
but none of the individual values, and in which no group of entities can
collude to decrypt any other individual's values. 
A related question, which would provide a solution to the previous problem,
is how to construct $n$-tuples of pseudorandom functions
$\{f, g_1, \dots g_{n-1}\}$ such that $f(i) = g_1(i) + \cdots +  g_n(i)$ 
for all positive integers $i$
and where the ability to compute any one of the functions does not
imply the ability to compute any of the other functions. 

The current implementation does not use a third party provider,
but its structure means that commodity cloud services could be used
to compute on and store sensitive data. All untrusted components 
can be pushed to federated or external resource providers, which enables 
scaling to large organizations. 
Computation of encrypted sums is 
easily parallelized, so can be spread across different cloud nodes,
or threads. Computation of the
pad sum by the client is also easily parallelized to different threads.

 
\section{Conclusions}
\label{sec:concl}


We defined the requirements for \name security:
that an individual has full access to
her own data, and may obtain help from the third party to analyze it, 
that individuals cannot access each other's data unless
they explicitly share privileges, 
that the third party learns nothing about the data values, 
and that some users can obtain statistics about a group of
individuals with help from the third party but learn nothing
more about the data values beyond what can be deduced from the 
statistic.
Such trust structures exist in many settings beyond presence systems, 
such as user studies, medical studies,
and usage data from social networking sites. 
Our family of simple, non-interactive \name protocols 
provides controls that users of \myU requested, 
and that are widely applicable in settings where there is a presumption of 
privacy or individuals have the power to opt out of data collection. 
As our implementation shows, our protocols are practical. 



\bibliographystyle{IEEEtran}
\bibliography{crypto,aware} 

\end{document}